\documentclass[prb,aps,twocolumn,showpacs]{revtex4-1}
\usepackage{dcolumn}
\usepackage{bm}

\usepackage[colorlinks,hyperindex, urlcolor=blue, linkcolor=blue,citecolor=black, linkbordercolor={.7 .8 .8}]{hyperref}
\usepackage{graphicx}
\usepackage{amsfonts}
\usepackage{amsmath}
\usepackage{amssymb}
\usepackage{amsbsy}
\usepackage{nicefrac}

\newcommand{\bpar}{$B_{\parallel}$}
\newcommand{\bparc}{$B_{\parallel}^c$}
\newcommand{\bperp}{$B_{\perp}$}
\newcommand{\as}{A$\rm_s$}
\newcommand{\at}{A$\rm_t$}
\newcommand{\ab}{A$\rm_b$}
\newcommand{\bs}{B$\rm_s$}
\newcommand{\bt}{B$\rm_t$}
\newcommand{\bb}{B$\rm_b$}
\newcommand{\cs}{C$\rm_s$}
\newcommand{\ct}{C$\rm_t$}
\newcommand{\cb}{C$\rm_b$}
\newcommand{\ds}{D$\rm_s$}
\newcommand{\ha}{H$_1$}
\newcommand{\hb}{H$_2$}
\newcommand{\hc}{H$_3$}
\newcommand{\hard}{$\langle1\bar{1}0\rangle$}
\newcommand{\easy}{$\langle110\rangle$}
\newcommand{\rxx}{$R_{xx}$}
\newcommand{\ryy}{$R_{yy}$}
\newcommand{\de}{$\Delta E_{1,0}$}

\begin{document}

\title {Heterostructure Symmetry and the Orientation of the Quantum Hall Nematic Phases}

\author{J. Pollanen}
\affiliation{Institute of Quantum Information and Matter, Department of Physics, California Institute of Technology, 1200 E. California Blvd., Pasadena, California 91125}
\author{K.B. Cooper}
\altaffiliation[Now at ]{Jet Propulsion Laboratory, Pasadena, CA 91109.  This work done while at the California Institute of Technology.}
\affiliation{Institute of Quantum Information and Matter, Department of Physics, California Institute of Technology, 1200 E. California Blvd., Pasadena, California 91125}
\author{S. Brandsen}
\affiliation{Institute of Quantum Information and Matter, Department of Physics, California Institute of Technology, 1200 E. California Blvd., Pasadena, California 91125}
\author{J.P. Eisenstein}
\affiliation{Institute of Quantum Information and Matter, Department of Physics, California Institute of Technology, 1200 E. California Blvd., Pasadena, California 91125}
\author{L.N. Pfeiffer}
\affiliation{Department of Electrical Engineering, Princeton University, Princeton, NJ 08544}
\author{K.W. West}
\affiliation{Department of Electrical Engineering, Princeton University, Princeton, NJ 08544}

\date{\today}

\begin{abstract}
Clean two-dimensional electron systems in GaAs/AlGaAs heterostructures exhibit anisotropic collective phases, the quantum Hall nematics, at high Landau level occupancy and low temperatures.  An as yet unknown native symmetry-breaking potential consistently orients these phases relative to the crystalline axes of the host material.  Here we report an extensive set of measurements examining the role of the structural symmetries of the heterostructure in determining the orientation of the nematics.  In single quantum well samples we find that neither the local symmetry of the confinement potential nor the distance between the electron system and the sample surface dictates the orientation of the nematic.  In remarkable contrast, for two-dimensional electrons confined at a single heterointerface between GaAs and AlGaAs, the nematic orientation depends on the depth of the two-dimensional electron system beneath the sample surface.
\end{abstract}


\maketitle

\section{Introduction}
High mobility two dimensional electron systems (2DES) in GaAs/AlGaAs heterostructures have provided some of the earliest evidence for the existence of electronic nematic liquid crystals \cite{fradkin10}.  In such a nematic, electron-electron interactions can stabilize a phase with translational invariance but spontaneous orientational order.  If the orientational order is pinned by a weak symmetry-breaking field (arising from the host crystal structure, or applied externally), the resistivity of the electron system becomes anisotropic, a readily observable signature.  That such phases can arise out of a collection of point-like electrons, in contrast to a conventional nematic liquid crystal composed of elongated molecules, is remarkable.

Electrical transport experiments \cite{lilly99a,du99} on ultra-clean GaAs-based 2DESs revealed that a strong anisotropy in the longitudinal resistivity develops at temperatures below about $T \sim 150$ mK when a perpendicular magnetic field \bperp\ has positioned the Fermi level near half filling of the second (or a few higher) excited Landau level (LL).  This anisotropy contrasts sharply with the observed isotropy of the resistivity, in the same samples, at half filling of the ground and first-excited LL, where electron-electron interactions lead to strongly correlated phases lacking orientational order, and at zero and low magnetic field where interactions are generally weak and the 2DES behaves semi-classically.  These early observations are in qualitative agreement with prior theoretical work \cite{koulakov96a,koulakov96b,moessner96}, at the Hartree-Fock level, which predicted the existence of charge density wave (or ``stripe'') phases at half-filling of the same LLs as found in experiment.  Additional theoretical work, incorporating quantum and thermal fluctuations, led to the prediction that static long-range stripe order was likely absent, being replaced, via a Kosterlitz-Thouless transition, by nematic orientational order at low temperatures \cite{fradkin99,fradkin00}.  There is both experimental and additional theoretical support for this scenario \cite{cooper02,wexler01,fogler02}.

An important aspect of the experimental observations of the nematic phases in 2DESs that is not understood is their consistent orientation relative to the crystalline axes of the host GaAs lattice.  In all but a very few cases \cite{zhu02,twosubbands}, the observed transport anisotropy is oriented such that the measured resistivity is large when the mean direction of the current flow is parallel to the \hard\ crystal axis and small when the current flow is parallel to \easy, thus suggesting that the stripes align parallel to \easy.  Since the crystalline symmetry of bulk GaAs provides no distinction between the orthogonal \hard\ and \easy\ directions, the experimental observation of just such a distinction must arise from the fact that the 2DES resides not in pristine bulk GaAs, but instead in a complex semiconductor heterostructure grown one atomic layer at a time.  In particular, the absence of mirror symmetry across the 2DES plane in such heterostructures opens the door to a distinction between \hard\ and \easy, even if it does not identify the mechanism whereby the nematic phases sense that distinction \cite{kroemer99}.

In this paper we report on how the nematic phases respond to two types of controlled modification of heterostructure symmetry.  In the first, we examine the nematic phases in a set of samples in which the 2DES is confined to a single quantum well that is either doped symmetrically or asymmetrically.  In this way we show that in these samples the sign of the local perpendicular electric field experienced by the 2D electrons does not alter the orientation of the nematic phases.  In the second set of experiments, the distance between the 2DES and the surface of the heterostructure is the controlled variable.  Here we find that if the 2DES is confined in a quantum well, the orientation of the nematic phases in our samples is unaffected by the distance to the surface.  In contrast, when the 2DES is confined at a single heterointerface, we find that the stripes lie along the ``normal'' direction ({\it i.e.} parallel to \easy), when the surface is fairly close to the heterointerface, but along \hard\ when the surface is more remote.  We discuss these results in the context of the various symmetry-breaking mechanisms that have been proposed theoretically to be responsible for the orientation of the quantum Hall nematic phases.

\section{Experimental}
\subsection{Samples}
The samples used in this work are conventional modulation-doped GaAs/AlGaAs heterostructures grown by molecular beam epitaxy (MBE) on $\langle 001 \rangle$-oriented GaAs substrates.  A total of 13 independently grown samples have been examined.  Ten of these are single GaAs quantum wells embedded in the alloy Al$_{0.2}$Ga$_{0.8}$As.  Silicon delta-doping layers in the alloy are set back a distance $d_t$ above and $d_b$ below the quantum well and populate it with a high mobility two-dimensional electron system.  These quantum well samples comprise three groups, A, B, and C, of three samples each, (\as, \at, \ab; \bs, \bt, \bb; and \cs, \cb, \ct), plus one additional sample \ds.  For samples \as, \bs, \cs, and \ds, the doping is symmetric, {\it i.e.} $d_t=d_s$, while for samples \at, \bt, and \ct\  (\ab, \bb, and \cb) the doping is asymmetric, with $d_b>d_t$ ($d_b<d_t$).  The doping asymmetry $d_t$:$d_b$ is 1:4 (4:1) in samples \at\ and \ct\ (\ab\ and \cb) and 1:1.7 (1.7:1) in sample \bt\ (\bb).  As intended, the measured \cite{meas} electron density in the 2DES in the group A and group C samples, and in sample \ds, are nearly identical, ranging from $n=2.7$ to $2.9 \times 10^{11}$ cm$^{-2}$.  The density in the three group B samples ranges from $n = 3.0$ to $3.1 \times 10^{11}$ cm$^{-2}$.  The width of the quantum well is $d_w=30$ nm for the group A and C samples, 32 nm for the group B samples, and 28.3 nm for sample \ds.

In addition to these quantum well samples, three single heterointerface samples, \ha, \hb, and \hc, were examined.  In these, the 2DES resides on the GaAs side of an interface between a GaAs and Al$_{0.32}$Ga$_{0.68}$.  A single Si delta-doping layer in the AlGaAs is positioned $d_t= 80$ nm above the interface.  The measured \cite{meas} 2DES density in these three samples ranges from $n=2.0$ to $2.3 \times 10^{11}$ cm$^{-2}$.

For all six group A and B samples, and for samples \ds\ and \ha, the distance $d_{cap}$ between the upper Si delta doping layer at the sample top surface is $d_{cap}=110$ nm,  For the three group C samples and sample \hb, this ``cap layer'' thickness was increased ten-fold to $d_{cap}=1010$ nm. Finally, for sample \hc, $d_{cap}=2010$ nm.  In none of the samples was an additional doping sheet placed within the cap layer.

Although the low temperature electron mobility among the 13 samples varied from 6 to $20 \times 10^{6}$ cm$^2$/Vs, all exhibited robust transport signatures of the fractional quantized Hall effect and, most importantly, the quantum Hall nematic phases at $\nu = nh/eB_{\perp} = 9/2$ and 11/2 filling factor.  (Since the degeneracy of a single, spin-resolved LL is $eB_{\perp}/h$, $\nu=9/2$ corresponds to half-filling of the lower spin branch of the $N=2$ second excited LL \cite{groupb}.) The important parameters for the various samples are summarized in Table I.  
\begin{table}
\begin{center}
\begin{tabular}{cccccccccc}
\hline
\hline
Sample & Type & & $n$ & & $d_{cap}$ & & $d_t$ & & $d_b$\\
\hline
\as\ & 30 nm QW & & 2.7 & & 110 & & 106 & & 106\\
\ab\ & '' & & 2.8 & & 110 & & 265 & & 66\\
\at\ & " & & 2.7 & & 110 & & 66 & & 265\\
\hline
\bs\ & 32 nm QW & & 3.0 & & 110 & & 98 & & 98\\
\bb\ & " & & 3.0 & & 110 & & 135 & & 78\\
\bt\ & " & & 3.1 & & 110 & & 78 & & 135\\
\hline
\cs\ & 30 nm QW & & 2.9 & & 1010 & & 106 & & 106\\
\cb\ & " & & 2.8 & & 1010 & & 265 & & 66\\
\ct\ & " & & 2.7 & & 1010 & & 66 & & 265\\
\hline
\ds\ & 28.3 nm QW & & 2.8 & & 110 & & 106 & & 106\\
\hline
\ha\ & SI & & 2.2 & & 110 & & 80 & & n/a\\
\hb\ & SI & & 2.3 & & 1010 & & 80 & & n/a\\
\hc\ & SI & & 2.0 & & 2010 & & 80 & & n/a\\
\hline\hline
\end{tabular}
\caption{Structural parameters of all GaAs/AlGaAs quantum well (QW) and single heterointerface (SI) samples used in this work. 2DES densities $n$ in units of $10^{11}$ cm$^{-2}$, cap layer thicknesses $d_{cap}$ and doping setback distances $d_t$ and $d_b$ in nm.}
\end{center}
\end{table}

\subsection{Transport measurements}
Each sample consists of a $\sim5 \times 5$ mm square chip cleaved from its parent MBE wafer. The crystallographic orientation of each sample is unambiguously established by one or more methods, including visual inspection of known surface defects \cite{cooper03}. The samples are mounted on a rotation stage and thermally anchored to the mixing chamber of a dilution refrigerator.  The rotation stage allows the magnetic field supplied by a superconducting solenoid to have components both perpendicular (\bperp) and, when needed, parallel (\bpar) to the 2DES plane.  The in-plane field \bpar\ is directed along the \easy\ crystallographic direction.  

Eight diffused In (or In-Sn) ohmic contacts positioned at the corners and side midpoints of each sample enabled standard low frequency ac electrical transport measurements.  Excitation currents were kept small enough (typically 10 nA) to avoid electron heating.

The measurements reported here focus on the longitudinal resistances \rxx\ and \ryy\ of the nematic phase at $\nu = 9/2$, with the $\hat{x}$ and $\hat{y}$ directions corresponding to the \hard\ and \easy\ crystallographic axes, respectively. For these measurements, two opposing side midpoint ohmic contacts are used to inject and withdraw current, while the voltage difference between the two corner contacts on one side of the mean current flow axis is recorded.  Typically, \rxx\ and \ryy\ can differ by factors of order a few, even at high temperatures and at zero magnetic field.  We attribute this to contact misalignments, inhomogeneities in the 2DES density, and other mundane sources.  In any case, this effect in no way interferes with the identification of the nematic phases since the latter exhibit vastly larger differences between \rxx\ and \ryy\ that are highly temperature and filling factor dependent.

\subsection{Signatures of the Quantum Hall Nematic}

Figure 1 illustrates the basic phenomenology of the quantum Hall nematic phases.  In Fig. 1a the measured longitudinal resistances \rxx\ and \ryy\ are compared, at $T=50$ mK, in sample \as, a symmetrically doped 30 nm quantum well.  The resistances are plotted versus perpendicular magnetic field \bperp\ (here the sample is not tilted, hence \bpar\ = 0) over a range encompassing both the nematic phase at $\nu = 9/2$ in the $N=2$ LL and the incompressible fractional quantized Hall state at $\nu=7/2$ in the $N=1$ first excited LL.  At $\nu =9/2$, near \bperp\ = 2.5 T, \rxx\ is more than 1000 times larger than \ryy.  Around $\nu = 7/2$ \rxx\ and \ryy\ differ somewhat, but by a far smaller factor than at $\nu = 9/2$.

\begin{figure}
\begin{center}
\includegraphics[width=1 \columnwidth]{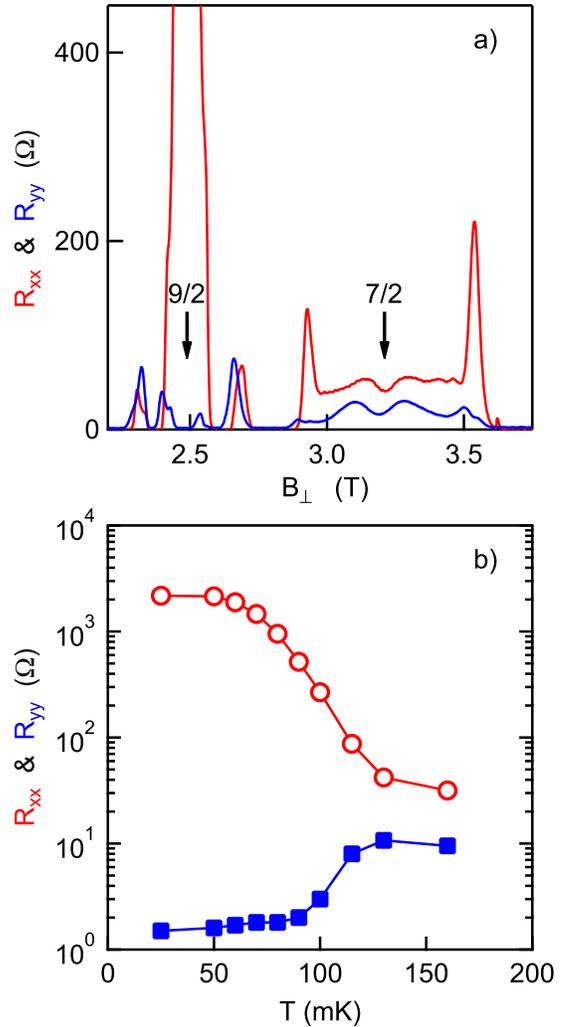}
\end{center}
\caption{(Color online) Basic transport signature of the quantum Hall nematic phases. a) Resistances measured along the \hard\ ($R_{xx}$, red) and \easy\ ($R_{yy}$, blue) crystal directions {\it vs.} perpendicular magnetic field at $T=50$ mK in sample \as. A giant anisotropy in the resistance is apparent at $\nu = 9/2$.  b) Temperature dependences of the measured resistances at $\nu = 9/2$.}
\end{figure}
The temperature evolution of the resistive anisotropy at $\nu = 9/2$ is displayed in Fig. 1b.  Above $T\sim 130$ mK \rxx\ and \ryy\ are only weakly temperature dependent, with \rxx\ exceeding \ryy\ by about a factor of 3, comparable to the difference seen around $\nu = 7/2$.  In contrast, as the temperature is reduced \rxx\ grows rapidly while \ryy\ falls.  This rapid onset of extreme anisotropy is characteristic of the nematic phases at half-filling of the $N\ge2$ Landau levels.  No analogous effect is seen at $\nu = 7/2$ or $\nu = 5/2$ in the $N=1$ LL or at $\nu = 3/2$ or $\nu = 1/2$ in the $N=0$ lowest LL \cite{N1}.
 
As reported earlier \cite{pan99,lilly99b}, a relatively small in-plane field, directed along the \easy\ axis, is sufficient to interchange the hard and easy transport directions of the nematic phases.  Theoretical analysis \cite{jungwirth99,stanescu00} has shown that the in-plane magnetic field, through its mixing of Landau levels and quantum well subbands, creates an extrinsic rotational symmetry breaking potential that competes with the native symmetry breaker responsible for the orientation of the resistive anisotropy at \bpar=0.  Indeed, the ``critical'' in-plane magnetic field, \bparc, needed to reorient the anisotropy provides a measure of the strength of the native symmetry breaking potential; $\sim$1 mK per electron being typical.  However, as we explain below, difficulties arise when comparing \bparc\ among samples with differing structures.

\section{Results}
\subsection{Quantum Well Symmetry}
To a good approximation, the density of electrons transferred from a Si delta-doping layer in an AlGaAs alloy across an interface to GaAs is simply $n = (\kappa \epsilon_0/e) \Delta V_c/d$, where $d$ is the distance between the doping layer and the interface, $\Delta V_c$ the conduction band edge offset at the interface, and $\kappa$ the alloy dielectric constant \cite{doping}.  The validity of this approximation relies on the dopant concentration being large enough to pin the Fermi level in the Si layer close to the conduction band edge.  We have found that it provides a highly successful design rule for a wide variety of 2D electron systems in GaAs-based heterostructures.

For a single quantum well doped from both sides, the donors create a perpendicular electric field at the location of the 2DES given approximately by $E_{\perp}^d=\Delta V_c (d_t-d_b)/2 d_b d_t$, where $d_t$ and $d_b$ are the doping setback distances.  Since in the same approximation the total electron density in the 2DES is given by $n=(\kappa \epsilon_0/e)\Delta V_c (d_t+d_b)/d_b d_t$ we have
\begin{equation}
E_{\perp}^d=n \frac{e}{2\kappa \epsilon_0}\frac{d_t-d_b}{d_t+d_b}.
\end{equation}
$E_{\perp}^d$ is a convenient quantitative measure of the asymmetry of the potential well confining the 2DES. Here and below, a {\it positive} electric field points along the MBE growth direction.

The nine quantum well samples in groups A, B, and C were designed specifically to determine whether confinement asymmetry was important in determining the orientation of the quantum Hall nematic phases.  Using Eq. 1 and the doping setbacks listed in Table I, the donor electric field at the 2DES is $E_{\perp}^d=+ (-) 1.2\times 10^6$ V/m for samples \ab\ and \cb\  (\at\ and \ct) and $E_{\perp}^d=+ (-)0.61\times 10^6$ V/m for sample \bb\ (\bt).  Obviously, $E_{\perp}^d=0$ for samples \as, \bs, and \cs.

Figure 2 shows \rxx\ and \ryy\ around $\nu = 9/2$ at $T=50$ mK in all six samples in groups A and B.  The large transport anisotropy characteristic of the nematic phase is clearly evident in all of the samples.  Furthermore, in all cases \rxx$\gg$\ryy\ at $\nu = 9/2$, demonstrating that the stripe orientation is the same in all six samples.  From these data we conclude that local symmetry of the potential confining the 2DES in these quantum well samples does not determine the orientation of the quantum Hall nematic phase at $\nu = 9/2$.  Though not shown in the figure, the same conclusion applies to the group C samples, and to the nematic phase at $\nu = 11/2$ in all nine samples. 
\begin{figure}
\begin{center}
\includegraphics[width=1 \columnwidth]{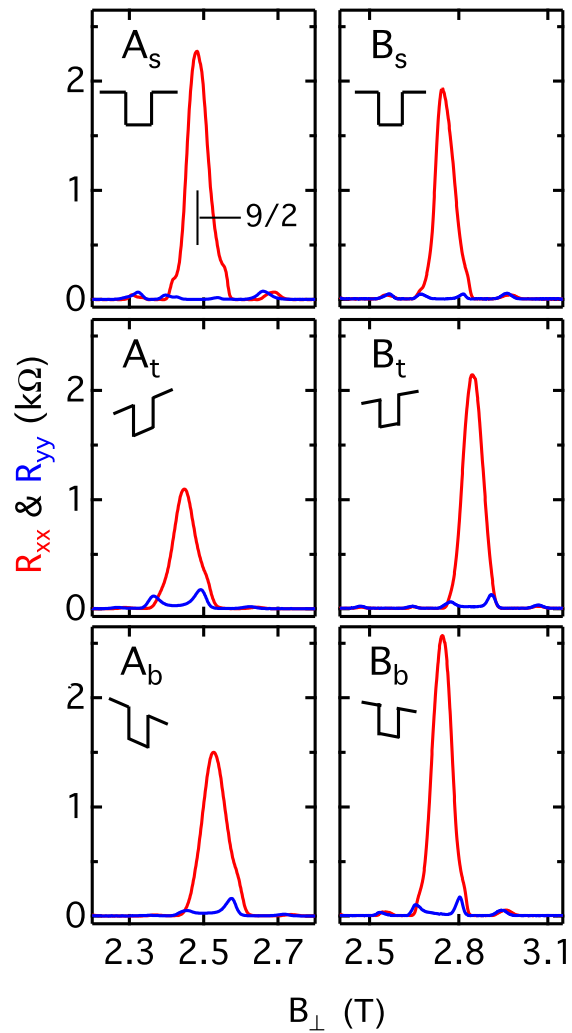}
\end{center}
\caption{(Color online) Resistances $R_{xx}$ (red) and $R_{yy}$ (blue) around $\nu = 9/2$ at $T = 50$ mK in all group A and B quantum well samples. In each case the quantum Hall nematic phase is robust and oriented such that the hard transport axis is \hard, irrespective of the symmetry of the quantum well (indicated by the sketches.) Though not shown, all group C samples exhibit the same orientation of the nematic phase.}
\end{figure}

\subsection{Cap Layer Thickness}
Dangling bonds at the physical surface of Ga-based heterostructures create mid-gap states which pin the Fermi level about $\Delta V_{s} \sim 800$ meV below the GaAs conduction band edge \cite{surface}.  In our samples the single Si delta-doping layer between the surface and the buried quantum well (or heterointerface) transfers electrons both to the quantum well {\em and} to the sample surface to bring the conduction band edge down to the Fermi level in the Si donor layer.  As a result, there is a perpendicular electric field, of magnitude $E_{\perp}^c= \Delta V_{s}/d_{cap}$, in the cap layer.  We note that this electric field does not affect the calculated donor electric field $E_{\perp}^d$ discussed in the previous section.

While all group A and group B quantum well samples have cap layers with thickness $d_{cap}=110$ nm, the three group C quantum well samples have a cap layer that is almost ten times larger: $d_{cap}=1010$ nm. For the group A and B samples $E_{\perp}^c \approx 9\times 10^6$ V/m, whereas for the group C samples $E_{\perp}^c \approx 1 \times 10^6$ V/m.  In spite of this large difference, the orientation of the quantum Hall nematic phase in the group C samples is the same as in groups A and B; the hard transport direction is \hard.  Apparently, the cap layer thickness does not affect the orientation of the nematic phases, at least in these quantum well samples.

\begin{figure}
\begin{center}
\includegraphics[width=1. \columnwidth]{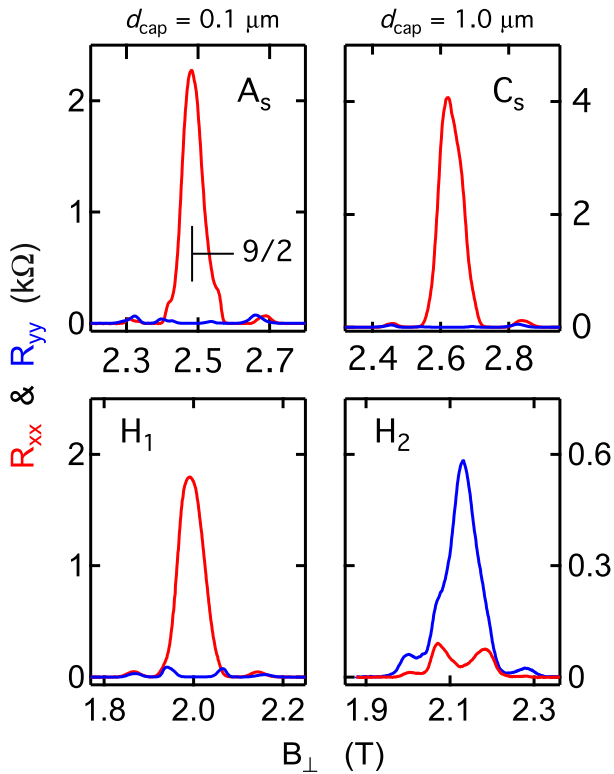}
\end{center}
\caption{(Color online) Contrasting effect of cap layer thickness in quantum well and single heterointerface samples.  Resistances $R_{xx}$ (red) and $R_{yy}$ (blue) around $\nu = 9/2$ at $T = 50$ mK.  The hard transport direction is along \hard\ in both the thin and thick cap quantum well samples \as\ and \cs, whereas it switches from \hard\ in the thin cap sample \ha\ to \easy\ in the thick cap sample \hb.}
\end{figure}
Remarkably, this conclusion does not carry over to the single heterointerface samples, \ha, \hb, and \hc, that we have examined. Figure 3 displays data from samples \ha\ and \hb.  For sample \ha, $d_{cap}=110$ nm, while for sample \hb, $d_{cap}=1010$ nm; these are the same thicknesses which distinguish the group A and B quantum wells from the group C samples.  Now we find that the transport hard axis at $\nu = 9/2$ is along the ``usual'' direction, $i.e.$ \hard, in sample \ha, but along \easy\ in sample \hb.  The same finding applies to the nematic phase at $\nu = 11/2$.  Since the 2DES density, doping setback $d_t$, and mobility in these two samples are nearly the same, it seems that the one structural difference between the two, the cap layer thickness $d_{cap}$, must be responsible for the differing orientation of the nematic phase.   Consistent with this, the third single heterointerface sample, H$_3$, for which $d_{cap}=2010$ nm, also exhibits the hard transport axis along \easy.  We emphasize that beyond these two thick cap heterointerface samples (\hb\ and H$_3$), there has, to our knowledge, been only one other reported example \cite{zhu02,twosubbands} of a 2D electron system in which the hard transport axis of the $\nu = 9/2$ nematic phase is along \easy\ instead of \hard\ (absent a symmetry-breaking in-plane magnetic field.)

\subsection{Effect of an In-Plane Magnetic Field}
As mentioned in section IIc, a small in-plane magnetic field \bpar\ directed along \easy\ is often sufficient to interchange the hard and easy transport directions of the quantum Hall nematic phases \cite{pan99,lilly99b,switch}.  This interchange occurs when the extrinsic symmetry-breaking potential due to the in-plane field overcomes the native potential responsible for orienting the nematic phase at \bpar=0.  The magnitude (and sign) of the extrinsic \bpar-induced potential is highly sensitive to the details of the quantum well confining the 2DES \cite{jungwirth99}.  

\begin{figure}
\begin{center}
\includegraphics[width=1 \columnwidth]{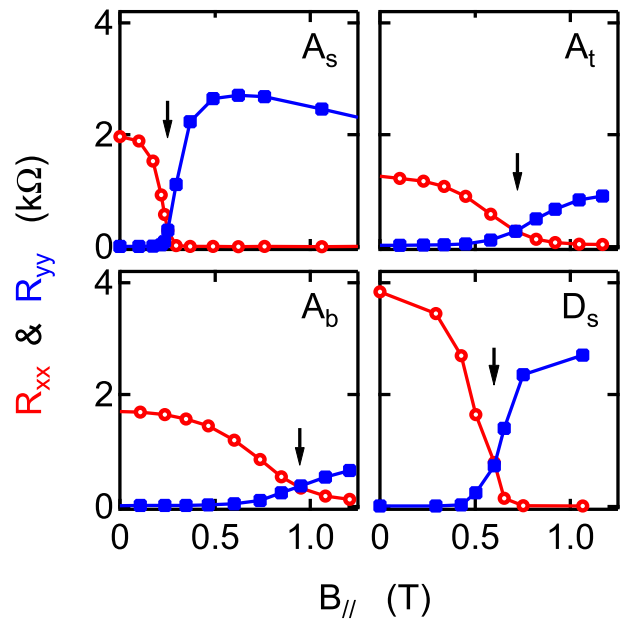}
\end{center}
\caption{(Color online) Interchange of resistance anistropy axes at $\nu = 9/2$ in samples \as, \at, \ab, and \ds\ due to an in-plane magnetic field \bpar\ applied along the \easy\ direction.  Arrows indicate ``critical'' in-plane field \bparc.  Data taken at $T=50$ mK.}
\end{figure}
Figure 4 compares the \bpar\ dependence of the resistive anisotropy in the $\nu = 9/2$ nematic phase in the symmetrically doped sample \as\ to that in its asymmetrically doped partner samples, \at, and \ab.  The figure reveals that the critical in-plane field \bparc\ is considerably larger in the asymmetric samples than in the symmetric sample.  It therefore seems that either: a) the native symmetry-breaking potential orienting the $\nu = 9/2$ nematic phase in samples \at\ and \ab\ is stronger than it is in sample \as, or b) the extrinsic symmetry-breaking potential due to the in-plane field is weaker in samples \at\ and \ab\ than it is in sample \as. 

The in-plane magnetic field creates its symmetry-breaking effect by mixing electric subbands of the confinement potential with the Landau levels arising from the perpendicular magnetic field \bperp.  The energy splitting \de\ between the ground and first excited electric subband is therefore crucial, with larger splittings leading to a weaker \bpar\ effect.  The calculated \cite{calc} values of this splitting are \de\ = 11.4 meV for the symmetric sample \as, and \de\ = 13.3 meV for the asymmetric samples \at\ and \ab.  To investigate whether this modest difference is sufficient to explain the large discrepancy in the \bparc\ values between sample \as\ and samples \at\ and \ab, an additional symmetric quantum well sample was grown and studied.  Structurally, sample \ds\ differs from sample \as\ only in the width of the quantum well: 28.3 nm vs. 30 nm, respectively.  The two samples have very nearly the same 2DES density and both clearly exhibit the quantum Hall nematic phases with the hard transport axis along \hard.  The calculated subband splitting in sample \ds\ is \de\ = 13.1 meV; as intended this is close to that in the two 30 nm asymmetric samples, \at\ and \ab.  As Fig. 4 shows, the interchange of the resistive anisotropy axes at $\nu = 9/2$ in sample \ds\ occurs at \bparc\ = 0.6 T.  This is much larger than the \bparc\  = 0.25 T seen in sample \as, and not far from the values found in samples \at\ and \ab.  These results demonstrate the sensitivity of the in-plane field effect to the details of the 2DES confinement potential and suggest caution in interpreting the magnitude of \bparc.

Figure 5 plots the observed \bparc\ values versus the calculated values of $\Delta E_{1,0}$ for all ten quantum well samples in this study \cite{heterocalc}.  The solid circles represent the symmetrically doped samples, the open circles those asymmetric quantum wells for which $d_b<d_t$, and the open diamonds the asymmetric wells for which $d_b>d_t$.  The symbols are blue for the group A samples, red for group B, black for group C, and green for sample \ds.  Not surprisingly, since the group A and C samples have the same doping profiles and nearly equal electron densities, their calculated subband splittings are very similar.  The symmetrically-doped samples \as\ and \cs\ exhibit almost identical critical fields, \bparc\ $\approx 0.25$ T.  The asymmetric samples \at, \ab, \ct, and \cb\ show considerable variation in the critical field (to which we return below) but, as already mentioned, \bparc\ is substantially larger than in the symmetric samples \as\ and \cs. On average, these asymmetric samples exhibit a \bparc\ close to that observed in sample \ds.  The group B samples, with their wider quantum wells (32 nm {\it vs.} 30 nm) and smaller doping asymmetries (1.7:1 {\it vs.} 4:1), have very nearly equal calculated subband splittings $\Delta E_{1,0}$ and measured critical fields \bparc.  Figure 4 demonstrates that, as expected, \bparc\ rises quickly with \de.

\begin{figure}
\begin{center}
\includegraphics[width=1 \columnwidth]{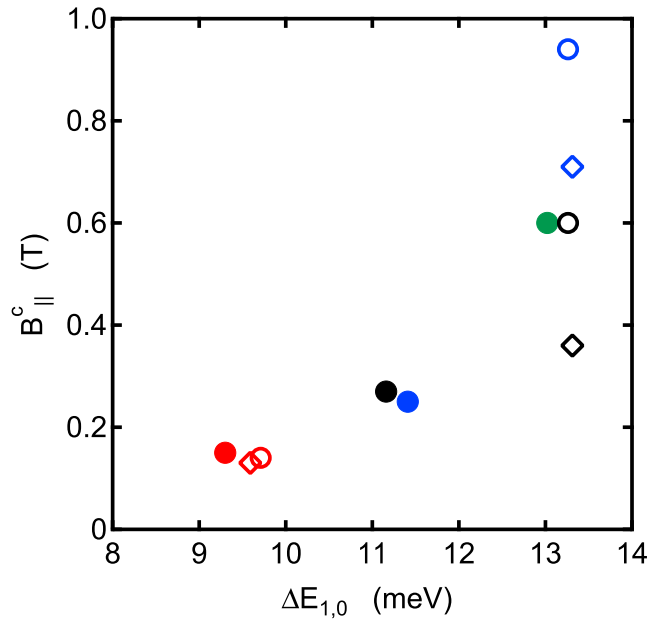}
\end{center}
\caption{(Color online) In-plane magnetic fields \bparc\ at which the resistive anisotropy axes interchange at $\nu = 9/2$ plotted versus the calculated energy splitting \de\ between the ground and first excited subbands of the quantum well confinement potential.  Solid dots: symmetrically doped samples. Open symbols: asymmetrically doped quantum wells, with $d_b>d_t$ (diamonds) and $d_b<d_t$ (circles). Blue, group A samples; red, group B; black, group C; green, sample \ds.}
\end{figure}

\section{Discussion}
There have been several suggestions for the origin of native symmetry-breaking field responsible for the consistent orientation of the quantum Hall nematic phases in 2D electron systems in GaAs/AlGaAs heterostructures.  Takhtamirov and Volkov \cite{takhtamirov00} and Rosenow and Scheidl \cite{rosenow01} suggested that anisotropy in the conduction band effective mass, due to the asymmetric confinement field in single heterointerfaces, was responsible. Fil \cite{fil00} pointed out that the piezoelectricity of GaAs explains why a one-dimensional charge density modulation in the 2DES prefers to lie along \hard\ {\it or} \easy, as opposed to, say, $\langle100\rangle$ \cite{cooper04}.  Subsequently, Fil \cite{fil01} argued that interference between the piezoelectric and deformation potential electron-phonon couplings lifts the degeneracy between \hard\ and \easy\ in a complicated way strongly dependent on the depth of the 2DES from the surface of the heterostructure.   More recently, Koduvayur {\it et al.} \cite{koduvayur11} have claimed that strain, in particular that arising from the electric field $E_{\perp}^c$ in the piezoelectric cap layer, explains the orientation of the nematic phases in both 2D electron and 2D hole systems \cite{tracy06}.  Finally, Sodemann and MacDonald \cite{sodemann13} have argued that a combination of Rashba and Dresselhaus spin-orbit interactions leads to a native symmetry-breaking potential which orients the nematic phases either along \hard\ or \easy\ depending upon the sign of the asymmetry of the 2DES confinement potential (which controls the sign of the Rashba interaction).  

The results shown in Fig. 2 clearly demonstrate that the symmetry of the confinement potential does not dominate the native symmetry-breaking potential responsible for orienting the quantum Hall nematic phases \cite{anomaly}.  This is consisent with an earlier report which compared the orientation of the $\nu = 9/2$ nematic in a symmetric quantum well sample with that found in a single heterointerface sample \cite{cooper01}.  Thus it appears that both the anisotropic effective mass models \cite{takhtamirov00,rosenow01} and the recent spin-orbit model \cite{sodemann13} are of at most secondary importance to the native symmetry-breaker.  

Interestingly, Fig. 5 shows that the critical field \bparc\ for interchanging the resistive anisotropy axes is larger in sample \ab\ than it is in sample \at, and larger in sample \cb\ than in sample \ct.  These four samples are asymmetrically doped quantum wells, with $d_b=d_t/4$ in samples \ab\ and \cb, but $d_b=4d_t$ in samples \at\ and \ct.  This might imply that the symmetry-breaking potential does contain a term proportional to the sign of the asymmetric donor electric field $E_{\perp}^d$. Although not strong enough to determine the orientation of the nematic, this term would influence the net strength of the symmetry-breaking potential, and thereby affect \bparc.  While appealing, we believe this conclusion is premature.  First, we note that the asymmetric quantum well samples \bb\ and \bt\ exhibit essentially identical \bparc\ values.  Although the donor electric fields in these group B samples are a factor of 2 less than those in the asymmetric group A and C samples ($E_{\perp}^d=\pm 0.6 \times 10^6$ {\it vs.} $\pm 1.2 \times 10^6$ V/m), a symmetry-breaking term proportional to $E_{\perp}^d$ should have shown up in samples \bb\ and \bt.  

Alternatively, the variation in \bparc\ values among the group A and C asymmetrically doped samples might be due to unintentional structural differences among these samples.  For example, the doping asymmetry of samples \ab\ (\cb) and \at\ (\ct) might not be simply opposite in sign, but also differ in magnitude.  Indeed, an ``upward'' diffusion of Si donors during  MBE growth would reduce the effective value of $d_b$ and increase that of $d_t$.  While our growth procedure is designed to minimize this effect, it is unlikely to be wholly absent.  In any case, such diffusion would increase $\Delta E_{1,0}$ in samples where $d_b<d_t$ and decrease it in samples where $d_b>d_t$.  Since \bparc\ increases with $\Delta E_{1,0}$, this diffusion mechanism might explain why \bparc\ is larger in sample \ab\ (\cb) than in sample \at\ (\ct). While the same diffusion effect would exist in the group B samples, it would have much less effect since the calculated subband splittings already show little sensitivity to the doping asymmetry.

We turn now to our observations regarding the cap layer thickness $d_{cap}$.   First, irrespective of its orientation, the quantum Hall nematic phase is clearly observed in samples \cs, \cb, \ct, \hb, and \hc, in all of which the 2DES lies more than 1 $\mu$m below the physical surface of the heterostructure \cite{dsurf}.   This is an order of magntitude larger than the period $\lambda$ of the charge density modulation of the ``stripe'' phases predicted by Hartree-Fock theory \cite{koulakov96a,koulakov96b,moessner96}.  For example, at $\nu = 9/2$, $\lambda \approx 6 \ell$, with $\ell=\sqrt{e\hbar/B_{\perp}}$ the magnetic length; $\lambda \approx 100$ nm in our samples. This observation thus discounts orientational symmetry-breaking mechanisms which rely on a nearby physical surface \cite{fil01}. 

Our findings also impact the recent suggestion \cite{koduvayur11} that piezoelectric strain, created by built-in electric fields in complex GaAs/AlGaAs heterostructures, governs the native symmetry-breaking potential.  For example, sample \as\ and \cs\ are both symmetrically doped 30 nm quantum wells.  Their 2DES densities, $n = 2.7$ and $2.9 \times 10^{11}$ cm$^{-2}$; subband splittings, \de\ = 11.4 and 11.2 meV; and critical in-plane fields, \bparc\ = 0.25 and 0.27 T; respectively, are all virtually identical.  Both exhibit strong quantum Hall nematic phases, with hard transport in the usual direction: \hard.  However, as mentioned in section III-B, the thickness and perpendicular electric field in the cap layer of these two samples differ by a factor of almost 10: $d_{cap} = 110$ nm and $E_{\perp}^c=9\times 10^6$ V/m in sample \as, but $d_{cap}=1010$ nm and $E_{\perp}^c=1\times 10^6$ V/m in sample \cs.  Consequently, the strain induced by the electric field $E_{\perp}^c$ in the cap layer is almost ten times smaller in sample \cs\ than it is in sample \as.  From this we conclude that piezoelectric strain in the cap layer due to Fermi level pinning at the surface is not the dominant source of the native symmetry-breaking potential in our quantum well samples.

In surprising contrast, the cap layer thickness strongly influences the symmetry-breaking potential in our single heterojunction samples. Both thick cap samples we examined, \hb\ and \hc, exhibited well-developed quantum Hall nematic phases with their hard transport direction along \easy.  These two samples, along with the unusual sample employed Zhu {\it et al.} \cite{zhu02}, are the only reported examples of 2D electron systems in which the quantum Hall nematic phases are oriented in this manner, at least in the absence of external perturbations or second subband occupation \cite{twosubbands}.  In their experiment, Zhu {\it et al.} employed a top-gated undoped single heterointerface.  In this density-tunable sample, the hard axis of the $\nu = 9/2$ nematic was found to switch from \hard\ to \easy\ when the density was increased beyond about $3 \times 10^{11}$ cm$^{-2}$.  Although the mechanism for this switching is unknown, it is interesting to note that the electric field $E_{\perp}^c$ in the cap layer of their sample was {\it negative} at all 2DES densities, with the switching observed for $E_{\perp}^c \lesssim -4 \times 10^{6}$ V/m.  In our heterointerface samples, the hard axis of the $\nu = 9/2$ nematic switched from \hard\ to \easy\ when $E_{\perp}^c$ was reduced from $+9 \times 10^6$ to $+1 \times 10^6$ V/m.  Hence, in both experiments the switching is observed as $E_{\perp}^c$ is changed in the same sense. The relationship, if any, between these observations remains to be investigated.

The single heterointerface results reported here suggest that the net native symmetry-breaking potential includes multiple contributions.  One of these could be piezoelectric strain due to the the electric field $E_{\perp}^c$ in the cap layer \cite{koduvayur11}.  This would be in competition with strain arising from the oppositely directed electric field $2E_{\perp}^d$ in the doping setback region between the Si donors and the heterointerface and the average electric field $E_{\perp}^d$ experienced by the 2D electrons themselves.  This last electric field might also contribute to the symmetry-breaking potential via the spin-orbit or effective mass anisotropy effects \cite{takhtamirov00,rosenow01,sodemann13} mentioned above. The strain due to the cap layer electric field might determine the orienation of the nematic phases in the thin cap heterointerface sample \ha, while these other effects win in the thick cap samples \hb\ and \hc.

While appealing, this scenario conflicts with our results in the quantum well samples.  For example, the two asymmetrically doped samples \at\ and \ct\ differ only in their cap layer thicknesses: 110 nm {\it vs.} 1010 nm, respectively.  Their large doping asymmetries result in electric fields in the quantum well and the top doping setback region which are the same sign and close in magnitude to those in the single heterointerface samples \ha\ and \hb.  Nevertheless, in contrast to samples \ha\ and \hb, these two quantum wells both display nematic phases with hard transport axes along \hard. Furthermore, as already mentioned, the symmetric quantum well samples \as\ and \cs, in which the electric fields in the top and bottom doping setbacks cancel and the average field in the quantum well itself vanishes, show the same orientation for the nematic (and essentially the same critical magnetic field \bparc) in spite of their very different cap layer thicknesses.

The different dependence upon cap layer thickness of the single heterointerface and quantum well samples presents an intriguing puzzle.  It suggests, perhaps not surprisingly, that the material both above and below the 2DES plays a role in the net native symmetry-breaking potential experienced by the nematic phases.  In the single interface samples, the 2DES sits atop a thick (typically $1 \mu$m) layer of GaAs whereas in the quantum well samples there are relatively thick AlGaAs alloy layers both above and below the thin GaAs quantum well containing the 2DES.  In both kinds of samples, these structures are supported by a ``cleaning'' superlattice consisting of many alternating thin layers of GaAs and AlGaAs and a GaAs buffer layer directly atop the GaAs substrate.  Certainly, the net strain profile and detailed electronic structure in these two types of many-layered heterostructures differ in detail and could be quite challenging to model quantitatively \cite{latticemismatch}.  
   
\section{Conclusion}
We have here reported on an extensive set of experiments, involving 13 distinct samples, designed to elucidate the nature of the native symmetry-breaking potential which orients the quantum Hall nematic phases that emerge at low temperature and high Landau level occupancy in clean 2D electron systems in GaAs.  Although the fundamental origin of the symmetry-breaking potential has not been determined, our findings significantly constrain theoretical models of it.  In particular, we have found the orientation of the quantum Hall nematic to be insensitive to the sign of the perpendicular electric field $E_{\perp}^d$ at the location of the 2DES.  This electric field, through its coupling to the orbital and spin-orbital degrees of freedom of the 2DES, has been suggested as responsible for the orientation of the nematic phases \cite{takhtamirov00,rosenow01,sodemann13}. 

In addition to the local symmetry of the 2DES confinement potential, we have also examined the quantum Hall nematic phases in samples with varying distances \cite{dsurf} between the 2DES and the physical surface of the heterostructure.  Increasing the cap layer thickness $d_{cap}$ from 0.1 $\mu$m to 1.0 $\mu$m was observed to have no effect on the orientation of the nematic phases in our quantum well samples but, remarkably, to interchange the hard and easy transport directions in our single heterointerfaces. While these opposite findings demonstrate that the cap layer thickness is not a reliable predictor \cite{koduvayur11} of the orientation, they do prove that a nearby surface is not essential for the robust pinning of it \cite{fil01} and, at the same time, open an interesting avenue for future research.

Finally, we have shown that the re-orientation of the transport anisotropy axes of the nematic phases due to an applied in-plane magnetic field \bpar, while fundamentally well-understood \cite{jungwirth99,stanescu00}, is not well-suited as a quantitative tool for comparing the strength of the native symmetry-breaker among disparate samples.  In particular, our measurements have shown that even a small change in the energy splitting between subbands in the confinement potential can significantly alter the critical in-plane field \bparc\ needed to re-orient the nematic phase.

\begin{acknowledgements}
We thank Inti Sodemann, Allan MacDonald, Steve Kivelson, and Eduardo Fradkin for discussions.  The Caltech portion of this work was supported by NSF Grant DMR-0070890, DOE grant FG02-99ER45766 and the Institute for Quantum Information and Matter, an NSF Physics Frontiers Center with support of the Gordon and Betty Moore Foundation through Grant No. GBMF1250. The work at Princeton University was funded by the Gordon and Betty Moore Foundation through Grant GBMF 4420, and by the National Science Foundation MRSEC Grant 1420541.
\end{acknowledgements}


\begin{references}
\bibitem{fradkin10} E. Fradkin, S.A. Kivelson, M.J. Lawler, J.P. Eisenstein, and A.P. Mackenzie, Annu. Rev. Condens. Matter Phys. {\bf 1}, 153 (2010).
\bibitem{lilly99a} M. P. Lilly, K. B. Cooper, J. P. Eisenstein, L. N. Pfeiffer, and K. W. West, Phys. Rev. Lett. {\bf 82}, 394 (1999).
\bibitem{du99} R.R. Du {\it et al.}, Solid State Commun. {\bf 109}, 389 (1999).
\bibitem{koulakov96a} A.A. Koulakov, M.M. Fogler, and B.I. Shklovskii, Phys. Rev. Lett. {\bf 76}, 499 (1996).
\bibitem{koulakov96b} A.A. Koulakov, M.M. Fogler, and B.I. Shklovskii, Phys. Rev. B {\bf 54}, 1853 (1996).
\bibitem{moessner96} R. Moessner and J.T. Chalker, Phys. Rev. B {\bf 54}, 5006 (1996).
\bibitem{fradkin99} E. Fradkin and S.A. Kivelson, Phys. Rev. B {\bf 59}, 8065 (1999).
\bibitem{fradkin00} E. Fradkin, S.A. Kivelson, E. Manousakis, and K. Nho, Phys. Rev. Lett. {\bf 84}, 84 (1982).
\bibitem{cooper02} K. B. Cooper, M. P. Lilly, J. P. Eisenstein, L. N. Pfeiffer, and K. W. West, Phys. Rev. B {\bf 65}, 241313 (2002).
\bibitem{wexler01} C. Wexler and A.T. Dorsey, Phys. Rev. B {\bf 64}, 115312 (2001).
\bibitem{fogler02} M.M. Fogler, Int. J. Mod. Phys. B {\bf 16}, 2924 (2002).
\bibitem{zhu02} J. Zhu, W. Pan, H. L. Stormer, L. N. Pfeiffer, and K. W. West, Phys. Rev. Lett. {\bf 88}, 116803 (2002).
\bibitem{twosubbands} We restrict our attention here to 2D systems in which only the ground electric subband of the confinement potential is occupied by electrons.  Alternative orientations of the nematic phases have been reported in 2D systems in which the ground and first excited subbands of the confinement potential are occupied \cite{pan00,liu13}.
\bibitem{pan00} W. Pan, T. Jungwirth, H. L. Stormer, D. C. Tsui, A. H. MacDonald, S. M. Girvin, L. Smr{\u c}ka, L. N. Pfeiffer, K. W. Baldwin, and K. W. West, Phys. Rev. Lett. {\bf 85}, 3257 (2000).
\bibitem{liu13} Yang Liu, D. Kamburov, M. Shayegan, L. N. Pfeiffer, K. W. West, and K. W. Baldwin, Phys. Rev. B {\bf 87}, 075314 (2013).
\bibitem{kroemer99} H. Kroemer, arXiv:cond-mat/9901016.
\bibitem{meas}  At low temperatures and after transient illumination with red light during the cooldown from room temperature.
\bibitem{groupb} Calculations \cite{calc} show that in all samples save those in group B, only the ground electric subband is occupied by electrons.  For the group B samples, the calculations suggest that about 6\% of the total electron density resides in the second electric subband at zero magnetic field.  In spite of this, the clear presence of the quantum Hall nematic phases at $\nu = 9/2$ and 11/2 in these samples demonstrates that at these magnetic fields the Fermi level resides in the $N=2$ LL of the ground subband, not the $N=0$ LL of the second subband.  The second subband is thus unoccupied at these fields.
\bibitem{cooper03} K.B. Cooper, Ph.D. thesis, California Institute of Technology (2003).
\bibitem{N1} Anisotropy does develop at $\nu = 7/2$ and 5/2 in 2D electron systems when an in-plane magnetic field is applied \cite{pan99,lilly99b}.  
Also, anisotropy at these filling factors is observed in 2D {\it hole} systems even in a purely perpendicular magnetic field \cite{manfra07}.
\bibitem{pan99} W. Pan, R.R. Du, H. L. Stormer, D. C. Tsui, L. N. Pfeiffer, K. W. Baldwin, and K. W. West, Phys. Rev. Lett. {\bf 83}, 820 (1999).
\bibitem{lilly99b} M. P. Lilly, K. B. Cooper, J. P. Eisenstein, L. N. Pfeiffer, and K. W. West, Phys. Rev. Lett. {\bf 83}, 824 (1999).
\bibitem{manfra07} M. J. Manfra, R. de Picciotto, Z. Jiang, S. H. Simon, L. N. Pfeiffer, K. W. West, and A. M. Sergent, Phys. Rev. Lett. {\bf 98}, 206804 (2007).
\bibitem{jungwirth99} T. Jungwirth, A.H. MacDonald, L. Smr{\u c}ka, and S.M. Girvin, Phys. Rev. B {\bf 60}, 15574 (1999).
\bibitem{stanescu00} T. Stanescu, I. Martin, and P. Phillips, Phys. Rev. Lett. {\bf 84}, 1288 (2000).
\bibitem{doping} A more refined model includes subtracting the confinement and Fermi energy of the 2DES from $\Delta V_c$.
\bibitem{surface}  Although all our samples have a thin (10 nm) GaAs layer at the very top, the Si donors are in the AlGaAs alloy.  Thus the effective value of $\Delta V_s$ is actually close to 1 eV.
\bibitem{switch} Typically the in-plane field orients the resistive anisotropy so that the hard axis is parallel to \bpar.  However, at large \bpar\ and in 2D electron systems in which both the ground and first excited electric subband in the confinement potential are occupied with electrons, the hard axis can be parallel $\it or$ perpendicular to the in-plane field, depending on details \cite{jungwirth99,pan00}.
\bibitem{calc} The calculation self-consistently solves the Schrodinger and Poisson equations in the local density approximation.
\bibitem{heterocalc} Results for the heterointerface samples \ha, \hb, and \hc\ are not included in Fig. 5 because of significantly greater uncertainty in the calculated $\Delta E_{1,0}$ values.  This greater uncertainty arises primarily from the poorly known concentration of unintentional impurities in the thick GaAs layer atop which the 2DES resides in these samples.
\bibitem{takhtamirov00} E.E. Takhtamirov and V.A. Volkov, JETP Letters {\bf 71}, 422 (2000).
\bibitem{rosenow01} B. Rosenow and S. Scheidl, Int. J. Mod. Phys. B {\bf 15}, 1905 (2001). 
\bibitem{fil00} D.V. Fil, Low Temp. Phys. {\bf 26}, 581 (2000).
\bibitem{cooper04} Experimental evidence in support of this conclusion can be found in K.B. Cooper, J.P. Eisenstein, L.N. Pfeiffer, and K.W. West, Phys. Rev. Lett. {\bf 92}, 026806 (2004).
\bibitem{fil01} D.V. Fil, J. Phys. Condens. Matter {\bf 13}, 11633 (2001).
\bibitem{koduvayur11} S.P. Koduvayur, Y. Lyanda-Geller, S. Khlebnikov, G. Csathy, M.J. Manfra, L.N. Pfeiffer, K.W. West, and L.P. Rokhinson, Phys. Rev. Lett. {\bf 106}, 016804 (2011).
\bibitem{tracy06} A closely-related suggestion was made previously by L.A. Tracy, J.P. Eisenstein, M.P. Lilly, L.N. Pfeiffer, and K.W. West, in Solid State Commun. {\bf 137}, 150 (2006).
\bibitem{sodemann13} I. Sodemann and A.H. MacDonald, arXiv:1307.5489.
\bibitem{anomaly}  Liu {\it et al.} \cite{liu13} observed that the hard axis of the quantum Hall nematic at $\nu = 13/2$ could switch from \hard\ to \easy\ if the confining quantum well was rendered even slightly asymmetric (in either sense) by electrostatic gating.  The sample in question was a wide quantum well with two occupied subbands, and the switching was observed when the system was close to a Landau level crossing.
\bibitem{cooper01} K.B. Cooper, M.P. Lilly, J.P. Eisenstein, T. Jungwirth, L.N. Pfeiffer, and K.W. West, Solid State Commun. {\bf119}, 89 (2001).
\bibitem{latticemismatch} We remark that the lattice mismatch between GaAs and AlGaAs is not obviously irrelevant.  For example, in our quantum well samples, the GaAs well is flanked by Al$_{0.2}$Ga$_{0.8}$As layers. These presumably random \cite{random} alloys produce an isotropic tensile strain of approximately $\epsilon_{xx}=\epsilon_{yy} \approx 4\times 10^{-4}$ in the thin GaAs layer.  For comparison, in GaAs the biaxial piezoelectric strain arising from a perpendicular electric field of $1 \times 10^7$ V/m is only $\epsilon_{xx}=-\epsilon_{yy} \approx 3 \times 10^{-5}$.  If the alloys are not truly random, owing, for example, to anisotropic growth kinetics, then the lattice mismatch might play a role in orienting the nematic phases.
\bibitem{random}A.R. Smith, Kuo-Jen Chao, C. K. Shihb, K. A. Anselm, A. Srinivasan, and B. G. Streetman, Appl. Phys. Lett. {\bf 69}, 1214 (1996).
\bibitem{dsurf} The total distance between the 2DES and the sample surface is the sum $d_{cap}+d_t$ of the cap layer thickness and the top doping setback distance.

\end{references}
\end{document}